\documentclass{emulateapj}

\usepackage{color}

\newcommand{\alphab}{{\boldsymbol\alpha}}
\newcommand{\omegab}{{\boldsymbol\omega}}
\newcommand{\kepler}{{\it Kepler}}
\newcommand{\kep}{{\it Kepler}--36}
\newcommand{\bc}{}
\newcommand{\rearth}{\,R$_\oplus$}
\newcommand{\rbondi}{\,R$_{\rm Bondi}$} 
\newcommand{\mearth}{\,M$_\oplus$}
\newcommand{\yr}{\,yr}
\newcommand{\myr}{\,Myr}
\newcommand{\gyr}{\,Gyr}
\newcommand{\gcmc}{\,g~cm$^{-3}$}
\newcommand{\au}{\,AU}
\newcommand{\kel}{\,K}

\newcommand{\paperacknowledge}{We are grateful to the referee for a constructive report that improved this manuscript. JEO acknowledges support by NASA through Hubble Fellowship grant HST-HF2-51346.001-A awarded by the
Space Telescope Science Institute, which is operated by the
Association of Universities for Research in Astronomy, Inc.,
for NASA, under contract NAS 5-26555. TDM is supported by the \kepler\ Participating Scientist Program, under grant NNX14AE11G. We thank Josh Carter for providing the MCMC chains from the original TTV analysis. We are grateful to Nikhil Mahajan, Leslie Rogers \& Yanqin Wu for helpful discussions. }

\shorttitle{Initial conditions of {\it Kepler}-36's planets}
\shortauthors{Owen, J.~E.~\& Morton, T.~D.}

\usepackage{amsmath,amssymb}

\begin{document}


\title{The initial physical conditions of \kep\ $b$ \& $c$}


\author{James E. Owen\altaffilmark{1}}
\affil{Institute for Advanced Study, Einstein Drive, Princeton, NJ 08540, USA}
\email{jowen@ias.edu}
\author{Timothy. D. Morton}
\affil{Department of Astrophysical Sciences, Princeton University, 4 Ivy Lane, Peyton Hall, Princeton, NJ 08544, USA}
\email{tdm@astro.princeton.edu}
%
%


\altaffiltext{1}{Hubble Fellow}


\begin{abstract}
The \kep\ planetary system consists of two exoplanets at similar separations (0.115\,\&\,0.128\au), which have dramatically different densities. The inner planet has a density consistent with an Earth-like composition, while the outer planet is extremely low-density, such that it must contain a voluminous H/He envelope. Such a density difference would pose a problem for any formation mechanism if their current densities were representative of their composition at formation. However, both planets are at close enough separations to have undergone significant evaporation in the past. We constrain the core-mass, core composition, initial envelope-mass, and initial cooling-time of each planet using evaporation models conditioned on their present-day masses and radii, as inferred from \kepler\ photometry and transit timing analysis. The inner planet is consistent with being an evaporatively stripped core, while the outer planet has retained some of its initial envelope due to its higher core-mass. Therefore, both planets could have had a similar formation pathway, with the inner planet having an initial envelope-mass fraction of $\lesssim10\%$ and core-mass of $\sim4.4$\mearth, while the outer had an initial envelope-mass fraction of order 15-30\% and core-mass $\sim7.3$\mearth.  Finally, our results indicate that the outer planet had a long ($\gtrsim30$\myr) initial cooling-time, much longer than would naively be predicted from simple timescale arguments. The long initial cooling-time could be evidence for a dramatic early cooling episode such as the recently proposed ``boil-off'' process.       
\end{abstract}





\section{Introduction}
The \kepler\ mission has altered our understanding of exoplanets, by detecting thousands of exoplanet candidates \citep[e.g.,][]{Mullally15}. The majority of these exoplanets are close ($\lesssim0.3$\au) to their parent star and small ($\lesssim3$\rearth, \citealt{Howard10,Howard12,Morton14,Silburt2015}), with most stars containing at least one close-in small planet \citep[e.g.,][]{Fressin13}. 
{\bc Understanding their origin would be a significant advance towards a complete theory of planet formation. }

Mass measurements of these systems using transit timing variations \citep[TTVs, e.g.][]{Wu13,Hadden14} or radial velocity follow up \citep[e.g.,][]{Weiss14} suggest that a large fraction contain voluminous H/He envelopes, as evidenced by their extremely low densities $\lesssim1$\gcmc\ \citep{Wu13,Jontof14,Rogers15}. At such close separations high energy photons from the star can heat the upper atmospheres of the exoplanets to high enough temperatures that they can drive a hydrodynamic evaporative outflow \citep[e.g.,][]{Lammer03,Murray-Clay09,oj12,oa15}. For close-in, low-mass planets, evaporation is an extremely important evolutionary process \citep[e.g.,][]{oj12,Lopez12,ow13,Howe15} that can completely strip-off an initial H/He envelope \citep{ow13,Lopez13}. This led \citet{ow13} and \citet{Lopez13} to argue that all low-mass, close-in planets could have contained H/He envelopes when they were born. Approximately $\sim50\%$ of the \kepler\ population is consistent with having been completely stripped by evaporation\citep{ow13}. Therefore, the planet population has been dramatically sculpted by evaporation, hiding any imprints from the formation process. 

For low-mass planets, the mass (and hence escape temperature) is dominated by any solid-core. \citet{ow13,Lopez12,Lopez13} speculated that taking into account a planet's evaporative history could be used to place constrains on the planet's composition at birth (which would otherwise be degenerate, e.g. \citealt{Rogers10}, based purely on an instantaneous comparison).

One of the most intriguing systems and strongest tests of the evaporation scenario is the \kep\ system \citep{Carter12}, which contains two planets at similar orbital separations (in a 7:6 resonance). The inner planet (which we refer to as $b$ throughout this work) has a mass of $\sim4.3$\mearth, density of $\sim6.8$\gcmc\ and separation of 0.115\au, while the outer planet (which we refer to as $c$) has a mass of $\sim7.7$\mearth, density of $\sim0.8$\gcmc\ and separation of 0.128\au. Thus, $b$ is consistent with being a solid, rocky planet, while $c$ contains a significant H/He envelope \citep{Carter12}. No planet formation scenario can produce such a dichotomy in density at such close separations; therefore, some evolutionary process must have occurred in order to produce such a system after $\sim6$\gyr\ of evolution. \citet{Lopez13} studied whether $b\,{\rm\&}\,c$ could have had identical envelope-mass fractions at birth, yet match the current observed properties of the planets and found using a simple model of the evaporation with a constant mass-loss ``efficiency'' that both planets could have started with an initial envelope-mass fraction of $\sim20$\%. {\bc Giant impacts after the gas disc dispersed have also been suggested, \citep{Liu2015,Schlichting2015}, although they have not been tested in any detail.}       

While the \citet{Lopez13} study was a convincing demonstration that evaporation could explain the current compositional dichotomy, it used a simplistic model of evaporation and made guesses about the core-masses and densities. Furthermore, the initial model for $b$ had a radius of $\sim$10\rearth, which is unlikely to constitute a ``bound'' planet \citep{ow15} at early times\footnote{Assuming the upper atmosphere is at an equilibrium temperature of $\sim1000$\kel, then a planetary radius of $10$\rearth\ is $>0.25$\rbondi, meaning it will be unstable.} and as such the initial envelope-mass fraction of $b$ was likely overestimated. 

The aim of this work is to go beyond the initial study of \citet{Lopez13} and directly infer the initial physical conditions of both $b\,{\rm\&}\,c$ by using a detailed evaporation model that predicts present-day mass and radius as a function of core-mass, core composition, initial envelope-mass fraction, and initial cooling-time.  This model has been explicitly obtained from hydrodynamic calculations and accounts for the significant drop in the ``efficiency'' parameter as a planet is stripped completely \citep{ow13}.  This is the first study to use such models to quantitatively constrain the possible ``birth'' compositions of an exoplanet system, allowing for direct connection between the system's present-day observed properties and formation theories {\bc \citep[c.f. the forward-modelling approach of,][]{Jin2014}}.

\section{Model}
We follow \citet{Lopez13} and assume that both $b$ and $c$ remain on circular orbits at their current separations for their entire lifetimes\footnote{We note small changes in orbital separation due to mass-loss will make little difference, $(\Delta\,a<0.4\%)$, but may effect resonance \citep{Teyssandier15}}. We calculate the evolution of each planet independently, including evaporation and bolometric irradiation by the central star.  As we evaluate this evolution on grids of initial physical conditions, we are able to use the inferred posterior distribution of the planets' present-day properties calculated from the TTVs to constrain these initial conditions.  Below, we describe the evolution models in more detail as well as the hierarchical model we use for the statistical inference.


\subsection{Planetary Evolution}

We follow \citet{ow13,ow15} and use the {\sc mesa} stellar evolution code \citep{Paxton11,Paxton13} to calculate the evolution of the planet.  We include evaporation by using the \citet{oj12} evaporation rates, {\bc we do not explicitly include the recently proposed  ``boil-off'' phase here \citep{ow15}; although, we do present a case where we attempt to model its consequences.} Bolometric irradiation is included using the $F_*-\Sigma$ approach as described by \citet{Paxton13} and \citet{ow15}. Since \kep\ is an evolved star we must include stellar evolution to get the radius of $c$ correct as it is inflated by stellar irradiation late in its evolution. We use {\sc mesa} again to perform this calculation using the best fit parameters described by \citet{Carter12}. The evolution of the X-ray luminosity in terms of the star's bolometric luminosity is then described by the \citet{Jackson12} empirical fits. For both $b$ \& $c$ we vary the following initial conditions: core-mass, core composition, initial H/He envelope-mass fraction and initial cooling-time (defined as the Kelvin-Helmholtz timescale). We then evolve the planet model until the current age of the \kep\ system ($6.8$\gyr). Note that the uncertainties in the age of the \kep\ system is $\sim1$\gyr; however, since the mass-loss is completely dominated at early times, such an uncertainty in age makes little difference to our results.

\subsection{Statistical Formalism}
The goal of our analysis is to infer the initial conditions of $b$ and $c$ given their currently observed properties.  However, these present-day quantities are not known precisely; uncertainties in physical quantities inferred from TTVs are often highly correlated both for an individual planet and for planets in the same system. We wish to preserve this information in our analysis, as it is both more accurate and constraining\footnote{For example uncertainties in $b$ \& $c$ masses are tightly correlated, thus an initial composition not consistent with the evolution of one planet can exclude an initial composition for the other.} than assuming random, independent errors on the individual planet parameters.  This sort of inference, in which a model is conditioned on quantities of which the observations are themselves uncertain is known as multi-level or hierarchical inference \citep[see e.g.,][]{Hogg10,DFM14,Demory2014,MortonWinn14,WolfgangLopez15,Wolfgang15}.  
While hierarchical inference can be computationally demanding, it can be greatly simplified if posterior samples of the intermediate quantities (in this case, the masses and radii of the planets) have been previously calculated \citep{Hogg10}.  Fortunately, the fit to the properties of the \kep\ planets was performed within a Bayesian formalism and their posterior distributions were estimated from MCMC sampling by \citet{Carter12}, so we are able to take advantage of this simplification, as described below.  

As stated above our model consists of four parameters for each planet, which we denote by the vector ${\boldsymbol\alpha}$. The vector of model parameters fitted to the TTV signal is denoted by ${\boldsymbol\omega}$ and the data is denoted by $D$.  Therefore, we may write the likelihood function for our model parameters ${\boldsymbol\alpha}$, by a change of variables as:
\begin{equation}
\mathcal{L}\left(D |\alphab\right)=\int\!\!{\rm d}{\boldsymbol\omega}\,p\left(D|{\boldsymbol\omega}\right)p\left({\boldsymbol\omega}|{\boldsymbol\alpha}\right)\label{eqn:like1}
\end{equation}
The fit to the TTV signal contain many parameters which are either not relevant to our model (e.g., inclination or limb darkening parameters) or that we choose to fix in our evolution models for the sake of computational efficiency (e.g., planetary orbital separations and stellar mass and radius, which have uncertainties of $\sim1-4$\%). Thus, we can partition $\omegab$ into two components: $\omegab_R$ which represent those parameters which our relevant to our model and $\omegab_N$ which represent parameters which are not. 
Now our model maps a value of $\alphab$ into a single point in the $\omegab_R$ space ($\omegab_R^\alpha$), thus $p(\omegab|\alphab)$ can be represented by a delta function \citep[e.g.][]{Wolfgang15}. Therefore, $p(\omegab|\alphab)$ becomes:
\begin{equation}
p\left(\omegab|\alphab\right)=p_0\left(\omegab_N\right)\delta\left(\omegab_R-\omegab_R^\alpha\right)
\label{eqn:delta1}
\end{equation}
where $p_0(\omegab_N)$ are the priors on the $\omegab_N$ parameters chosen by \citet{Carter12}. Now from Bayes' theorem we can write:
\begin{equation}
p\left(D|{\boldsymbol\omega}\right)\propto\frac{p\left({\boldsymbol\omega}|D\right)}{p_0\left(\omegab\right)}=\frac{p\left({\boldsymbol\omega}|D\right)}{p_0\left(\omegab_N\right)p_0\left(\omegab_R\right)}
\label{eqn:data1}
\end{equation}
where in the final equality we have used the fact the priors on $\omegab_N$ and $\omegab_R$ are independent. Therefore, substituting Equations \ref{eqn:delta1} \& \ref{eqn:data1} into Equation \ref{eqn:like1} and integrating we find:
\begin{equation}
\mathcal{L}\left(D |\alphab\right)\propto\frac{p\left({\omegab_R^\alpha}|D\right)}{p_0\left(\omegab_R^\alpha\right)}\label{eqn:Like1}
\end{equation}
Finally, we note that the model parameters used by \citet{Carter12} consider mass and radius ratios, and we must transform the prior probability in Equation~\ref{eqn:Like1} to the original set of co-ordinates $\omegab_R'$ used by \citet{Carter12} by use of a Jacobian ($J=|d\omega_R^\alpha/d\omega_R'^\alpha|$) 
such that the likelihood function we evaluate is:
  \begin{equation}
\mathcal{L}\propto\frac{p\left({\omegab_R^\alpha}|D\right)J}{p_0\left(\omegab_R'^\alpha\right)}
\end{equation} 
The  function $p\left({\omegab_R}|D\right)$ can be estimated directly by kernel density estimation on the MCMC sample provided by \citet{Carter12}. The prior distribution $p_0(\omegab_R')$ used by \citet{Carter12} was uniform. We consider uniform priors on all parameters: flat for core-mass and core composition and log-flat for initial envelope-mass fraction and initial cooling-time, {\bc although we checked these choices are not driving our results}. In the case of the core composition we use a variable that tracts the percentage of ice or iron in a rock core where the range $[-1,0]$ corresponds to 100\% iron (-1) through to 100\% rock (0) and $[0,1]$ corresponding to 100\% rock (0) through to 100\% ice (1) following the \citet{Fortney2007} mass-radius relations. Therefore we require our core composition parameter to be in the range $[-1,1]$.  While formation models should be able to predict the initial cooling-time, at the moment they are unable to do with any certainty, {\bc as they are sensitive to the uncertain nebular conditions, disc lifetimes, core formation timescales and migration history.} Thus we conservatively set the initial cooling-time to be bounded between 1\myr\ -- the minimum protoplanetary disc lifetime \citep[e.g.][]{Mamajek2009,Owen2011}, and 6.8\gyr, the age of the \kep\ system. Finally, in order for the planet to be considered ``bound'' at zero time we restrict the initial radius to be $<0.1$\rbondi.      

We consider three scenarios related to specific insights about their possible formation. First, we adopt a conservative approach and assume that the two planets properties are unrelated (scenario I), giving 8 free parameters. Second (scenario II), since the cores of both planets are likely to be formed from the same population of planetesimals and embryos, we assume both the cores of $b$ \& $c$ have the same composition and finally (scenario III) we assume that both planets have the same composition and that both planets experienced a ``boil-off'' phase \citep{ow15} after formation. {\bc While not accounting for the mass-loss (our stated ``initial envelope-mass-fractions'' will still be post-boil-off values), we can attempt to account for the \textit{thermodynamic} state post-boil off,} which leaves the planetary radii close to $\sim0.1$\rbondi. Using Figures~4 {\bc \& 6} of \citet{ow15} we restrict the planet to have a radius between 0.08 and 0.1\rbondi at the start of the calculation, {\bc we note this significantly reduces the available envelope-mass and cooling timescale parameter space}. Scenario II \& III contain 7 free parameters. Since $b$ essentially contains no H/He envelope today scenario I \& II will only place limits on the initial envelope-mass fraction, whereas scenario III will place an actual constraint on the value. We use the {\texttt emcee} affine-invariant MCMC sampler \citep{ForemanMackey2013} to perform the parameter estimation under the three different scenarios.  

\section{Results}
\begin{figure*}
\centering
\includegraphics[width=\textwidth]{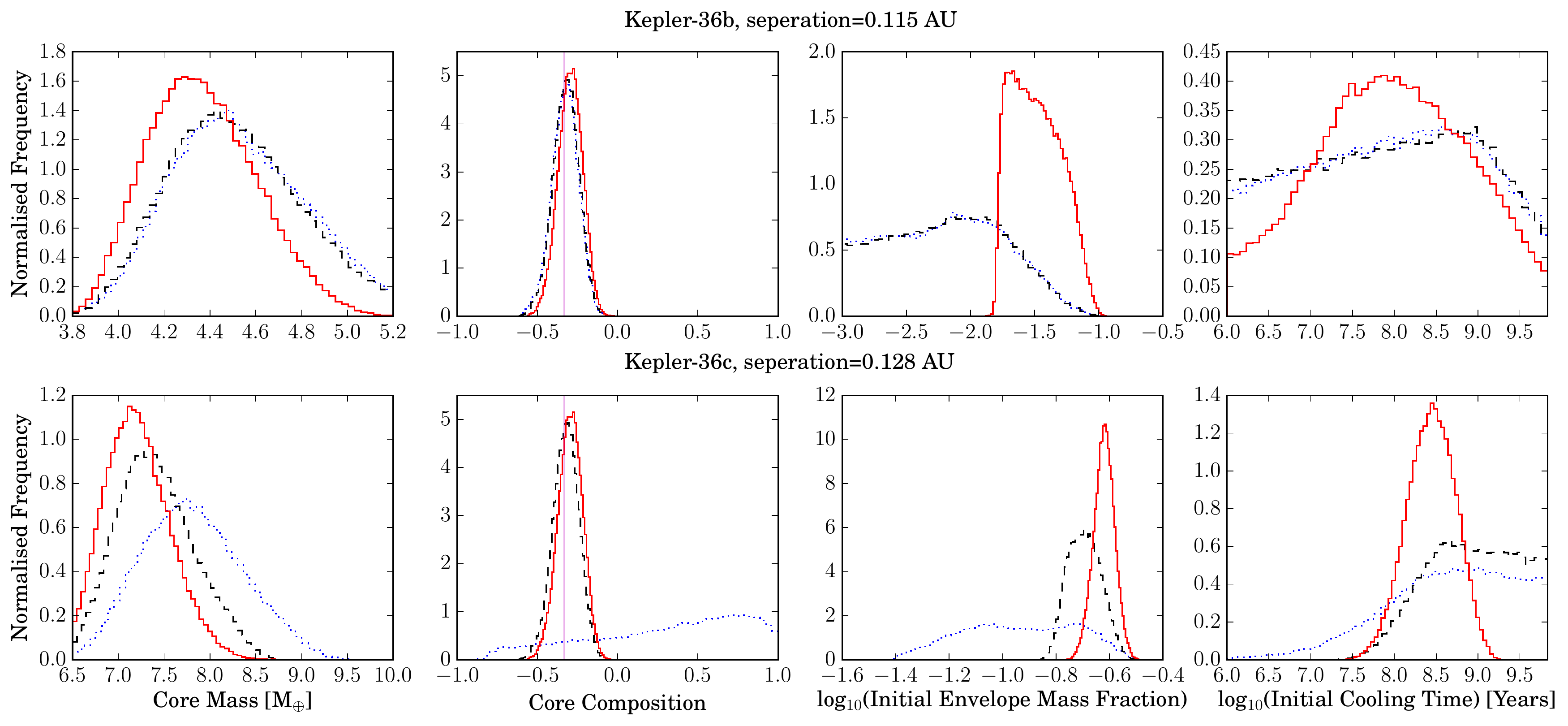}
\caption{Marginalised posterior distrubtions of all the planet properties for $b$ (top-row) and $c$ (bottom-row). Scenario I is shown as the dotted blue line, scenario II is the black-dashed line and scenario III is the red-solid line. The light-magenta line in the core composition panel shows an Earth-like composition \citep{Fortney2007}.}\label{fig:panel}
\end{figure*}

The results are shown in Figure~\ref{fig:panel}, where we show the posterior distributions of both $b$ \& $c$'s core-mass, core composition, initial envelope-mass fraction and initial cooling-times for our three scenarios (I: blue-dotted; II: black-dashed; III: red-solid). We get good constraints on most parameters, expect perhaps the initial cooling-time of $b$'s envelope in scenario I \& II and $c$'s core composition in scenario I. 

By linking the core composition of $b$ \& $c$ (scenarios II \& III)  we tighten the constraints on the parameters, especially the core composition and initial envelope-mass fraction of $c$. This is not surprising {\bc as  the present-day H/He envelope on $c$ prevents using its mass and radius to explicitly calculate a core composition} \citep{Carter12}.  We do rule out that $c$ has an exceedingly iron rich core.

As expected, the initial envelope-mass fraction for planet $b$ is only an upper limit in scenario I \& II, but not in scenario III when we require $b$ to have a significant atmosphere at early times.  Finally, we note the advantage of incorporating the full correlation structure of the measured quantities of both planets. The uncertainty in the core-mass and composition of planet $b$ is primarily driven by the directly measured uncertainties in planet mass and radius determined from the TTV analysis; in scenario III, by linking the planet parameters, we decrease these uncertainties. This is because some of the core compositions consistent with the measured mass and radius of planet $b$ are not consistent with the evolutionary history of planet $c$. 

\section{Discussion}

Our results place the first statistical constraints on the properties of any close-in low mass exoplanet at birth and indicate that current density dichotomy of the \kep\ system does not pose a problem with the idea that they both formed with significant H/He envelopes, consistent with the previous analysis of \citep{Lopez13}. In our most general case: scenario I, in which the cores of the planets are allowed to have completely independent composition is unlikely from a theoretical perspective. If they formed from the same population of planetesimals/embryos, the cores should have very similar compositions. Having the two planets form in vastly different regions of the disc (where the planetesimal compositions could in principle be different) then convergently migrate to their current location appears highly unlikely:  they are in a delicate 7:6 resonance that would require fine tuning to migrate into \citep{Quillen2013}.  Therefore, while presenting scenario I for completeness and illustrative purposes, we only discuss the implications of scenarios II and III further. 

\begin{figure*}
\centering
\includegraphics[width=\textwidth]{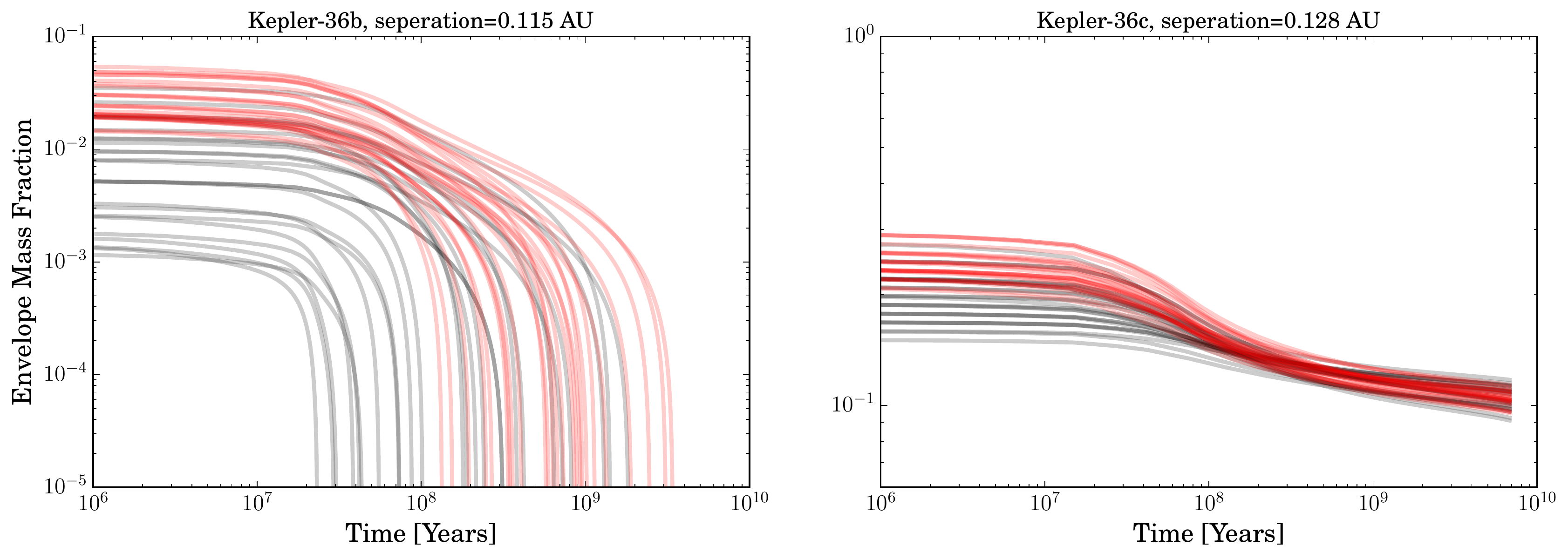}
\caption{The evolution of $b$'s (left) and $c$'s (right) H/He envelope-mass fraction. The black and red lines show 25 randomly selected evolutionary paths from the results for scenario II \& III respectively.}\label{fig:evolve}
\end{figure*}

The evolution of the envelope-mass fraction for $b$ \& $c$ are shown in Figure~\ref{fig:evolve}, where we randomly drawn cases from the results for scenario II (black) \& III (red).
These evolutionary curves show the role of core-mass in the evaporative history \citep[c.f.][]{Lopez13}. While $c$ retains a significant fraction of its original envelope to late times, the lower mass core of $b$ means that it can retain none of its original envelope. Therefore, by considering the evolution of planet $b$ we are able to conclude that $b$ contains no H/He envelope and is a naked core, compared to previous models which just considered the current mass and radius to put a constraint on the H/He envelope-mass fraction of $<0.4\%$ \citep{Carter12,Lopez2014}; {\bc whereas $c$'s initial and final envelope-mass is consistent with the single model calculated by \citet{Lopez13}}.    

Perhaps the most interesting result, robust across scenarios, is that we require $c$ to have a long initial cooling-time, considerably longer than the few Myr one would naively guess based on protoplanetary disc lifetimes and simple formation models {\bc \citep[e.g.][]{Rogers11,Lee2015}}. This is because models with small initial cooling-times have large initial radii, such that they lose more mass than those with longer cooling-times, therefore, requiring a larger initial envelope-mass to compensate. Eventually, such a large initial envelope-mass and short cooling-time means the planet envelope could no longer be hydrostatic and bound at early times, so the planet could not have had this initial structure. Such a long initial cooling timescale is a strong hint of a dramatic cooling process early in the planet's life. The ``boil-off'' process, a period of dramatic mass-loss and cooling just after disc dispersal \citep{ow15}, indeed produces this type of cooling and is the motivation for scenario III. {\bc It could also be hinting at evidence of inefficient heat transport \citep[e.g., semi-convection,][]{Garaud2013}, possibly occurring after giant impacts, \citep{Liu2015}.} One should be cautious with the results from just one planet, but this kind of modelling provides an avenue for evaluating whether the evolutionary histories of low-mass, close-in planets more generally may require a ``boil-off''--like process to have occurred.  

\subsection{The core-mass, envelope-mass plane}
\begin{figure}
\centering
\includegraphics[width=\columnwidth]{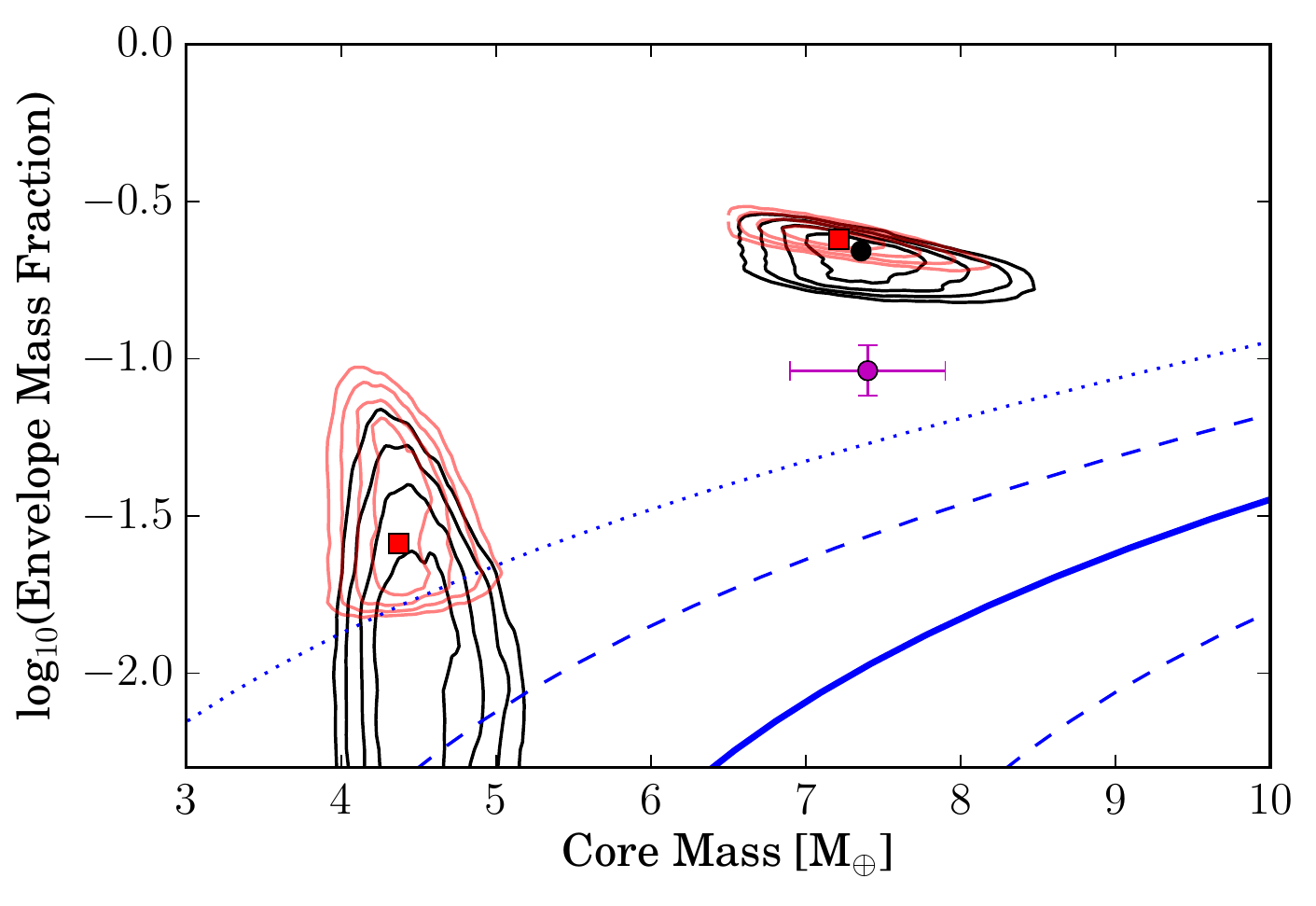}
\caption{Confidence intervals in the envelope-mass fraction--core-mass plane, shown as contours at 0.5-, 1.0-, 1.5-, 2.0-$\sigma$ for $b$ (left) and $c$ (right), in scenarios II (black) and III (red). {\bc The filled black points and red squares show} the best fit parameters. The magenta point shows the current composition of $c$ taken from our models. The blue lines show the observed relation (solid), 1.0- (dashed) and 2.0-$\sigma$ spread for the observed planet population today calculated from the \citep{Wolfgang15} mass-radius relationship.}\label{fig:X_mcore}
\end{figure}

One of the motivations of doing this work is we can use these models to peer at exoplanet structure shortly after formation. One test of a planet formation scenario is the kind of planetary compositions it produces and to first order this can be described in terms of a envelope-mass fraction--core-mass relation. Unfortunately, evaporation can dramatically change this relation when observed at late times \citep{ow13}. In Figure~\ref{fig:X_mcore} we show the envelope-mass-fraction, core-mass relation for $b$ \& $c$, as well as the general relation seen in the data at late times from \citep{Wolfgang15}, where we use \citealt{Lopez2014}'s mass, radius \& envelope-mass relations to convert Wolfgang et al.'s mass-radius relation to a core-mass, envelope-mass relation. 

Our results suggests that planets with larger core-masses accrete a larger initial envelope, which is not unexpected theoretically \citep[e.g.,][]{Rafikov2006,Lee2014,Lee2015}. {\bc The results are consistent within the spread of models found numerically by \citet{Mordasini2012}, and are consistent with the dusty and gas-poor scalings of \cite{Lee2015}}.  Since $c$ appears to be an outlier in terms of its current envelope-mass fraction, drawing inference regarding the more general nature of the {\it birth} envelope-mass, core-mass relation would be premature. However, the approach demonstrated in this work indicates that with, the now large and growing sample of well constrained low-mass exoplanets one can begin to trace out their initial properties. A well described birth envelope-mass, core-mass relation along with any intrinsic scatter would provide a strong constraint for any planet formation model.  

\section{Summary}

We have used exoplanet evolutionary models that include evaporation to statistically infer the properties of \kep\ $b$ \& $c$ shortly after formation. While $b$ \& $c$ have dissimilar densities today, they are consistent with having a common formation pathway, where the H/He envelope acquired by $b$ is completely lost due to evaporation, while the higher core-mass of $c$ allows it to retain a large fraction of its original envelope. Our results give promise that with a reasonable sample of well constrained planet masses and radii, we will be able to infer the birth envelope-mass--core-mass relation---similar to the more widely discussed planet mass--radius relation, but directly applicable to formation theories. The observations necessary to begin this exploration has already been provided by \kepler, and the upcoming TESS mission promises to deliver an even larger, possibly age-dependent sample.  Finally, $c$'s initial cooling-time is constrained to be long ($\gtrsim3\times10^{7}$\yr), significantly longer than predicted based on protoplanetary disc lifetimes of $<10$\myr; as such, the results indicate that $c$ may have undergone a dramatic cooling event early in its lifetime, such as the recently proposed ``boil-off'' process \citep{ow15}.    

\acknowledgements
\paperacknowledge 

\bibliographystyle{apj}



\end{document}